\documentclass[journal]{IEEEtran}

\usepackage{amsmath}
\usepackage{amsfonts}
\usepackage{graphics}
\usepackage{pstricks}
\usepackage{array}
\usepackage{float}
\usepackage{epsf}
\usepackage{graphicx}
\usepackage{multirow}
\usepackage{cite}
\usepackage{physics}
\usepackage{acronym}
\usepackage{comment}
\usepackage{upgreek}
\usepackage{subcaption}
\usepackage{nicefrac} 
\usepackage{amssymb}
\usepackage[font=footnotesize]{caption}
\usepackage{url}

\newcommand{\bfg}[1]{\boldsymbol{#1}}

\newcommand{\Ito}{It{\^o}'s}

\newcommand{\Wt}{Wiener process}

\newcommand{\OK}{\checkmark}

\acrodef{rocof}[RoCoF]{Rate of Change of Frequency}

\begin{document}

\title{Frequency Quality Assessment of GFM and GFL Converters and Synchronous Condensers}

\author{Taulant K{\"e}r{\c c}i,~\IEEEmembership{Senior Member,~IEEE},
  and Federico~Milano,~\IEEEmembership{Fellow,~IEEE}
  \thanks{T.~K{\"e}r{\c c}i is with Transmission System Operator,
    Innovation \& Planning Office, EirGrid, plc, Ireland.  F.~Milano
    is with School of Electrical and Electronic Engineering,
    University College Dublin, Belfield Campus, Dublin 4, Ireland.
    Corresponding author.  e-mail: federico.milano@ucd.ie.}
  \thanks{This work was partially supported by Sustainable Energy
    Authority of Ireland (SEAI) by funding Federico Milano through
    FRESLIPS project, Grant No.~RDD/00681.}
  \vspace{-6mm}
  }

\maketitle

\begin{abstract}
   This paper compares the impact of different conventional and emerging technologies and control strategies on frequency quality.  We study, in particular, the long-term dynamic performance of grid-forming (GFM) and grid-following (GFL) inverter-based resources (IBRs) as well as conventional synchronous machines.  
   Extensive simulations and several realistic scenarios consider both short-term and long-term aspects of frequency quality.  It is shown that, while overall GFM IBRs significantly improve frequency quality, a combination of GFL IBRs providing frequency support such as wind and batteries, and synchronous condensers, might be enough to meet similar frequency quality standards. 
   Another result of the paper is that the need for automatic generation control (AGC) becomes less clear in GFM IBR-dominated grids from a frequency quality perspective. 
\end{abstract}


\IEEEpeerreviewmaketitle




\vspace{-2mm}

\section{Introduction}

\subsection{Motivation}


Frequency quality refers to the ability of a power system to maintain frequency within secure/defined limits during normal and abnormal operating conditions \cite{entsoe}.  The integration of variable non-synchronous renewable energy sources (RES) coupled with increased market integration and dynamics is challenging the frequency quality management task leading in some instances to poor frequency quality \cite{10253411, 10688761}.  This is particularly the case of systems that do not actively utilize RES for frequency control or have limited number of battery energy storage systems (BESS) that provide excellent fast (e.g., $<$ 1s response time) primary frequency control (PFC).  For instance, despite the Continental European (CE) power system being one of the biggest interconnected systems in the world, in 2023, it exceeded for the first time the annual target of less than 15,000 minutes outside $\pm$50 mHz range (15,389 minutes, i.e., more than 10.5 days of cumulative time) \cite{entsoe}.  

A promising technology capable of addressing poor frequency quality is grid-forming control (GFM) of converters \cite{8879610, 10444681}.  However, there are still many open practical research questions on how GFM compares to other technologies such as grid-following (GFL) inverter-based resources (IBRs), namely, GFL BESS and wind, and synchronous generators and synchronous condensers.  This paper aims at filling this gap by bringing together industry and academic expertise.  For instance, while the results of the paper are based on an IEEE benchmark power system, they closely represent the dynamic behavior of real-world power systems such as the All-Island power system (AIPS) of Ireland.  This is done by considering representative realistic scenarios, simulation setups and frequency quality parameters.

\vspace{-1mm}
\subsection{Literature Review}

The vast majority of existing literature and, in particular, literature on GFM, focuses on short-term aspects of frequency quality.  In the short term, the focus is on frequency stability, which is evaluated based on a variety of metrics, for example, the rate of change of frequency (RoCoF); and frequency Nadir and Zenith.  Less attention is given to long-term aspects of frequency quality such as the standard deviation of the frequency, $\sigma_{f}$; and minutes outside relevant frequency range \cite{9714816, 9513281}.  
Reference \cite{11008671} assesses the short-term performance of different GFM control architectures, including GFM virtual synchronous machine (VSM) and droop, in dealing with frequency stability challenges.  
%
%
In the same vein, the authors in  \cite{jalali2023impact} demonstrate that GFM inverters enhance the frequency stability through their inertial response, but also by providing system strength to nearby IBRs and, thus, improving their fault ride through performance.  
Reference \cite{iet} shows that while GFM IBRs can replace synchronous machines, the type of control influences the frequency behavior, especially in the short-term response.  However, a common shortcoming of the aforementioned references is the lack of industry perspective.  

\vspace{-1mm}
\subsection{Contributions}

This industry-oriented paper addresses the above limitations and brings the following main novel contributions.
\begin{itemize}
\item A systematic and in-depth analysis and comparison of impact of GFM on frequency quality.  In particular, the impact of GFM on long-term frequency quality is, for the first time, thoroughly studied and discussed.
\item The paper shows that while GFM improves almost all aspects of frequency quality, similar standards could be achieved by a combination of GFL devices and synchronous machines such as GFL BESS and synchronous condensers, respectively.
\item The paper shows that the role of automatic generation control (AGC) for frequency quality reduces in GFM-dominated grids indicating a need to revisit its purpose.
\end{itemize}

\vspace{-1mm}
\subsection{Organization}

The paper is organized as follows.  Section~\ref{sec:model} presents the power system model based on stochastic differential algebraic equations.  Section~\ref{sec:case} discusses the results including for both normal and abnormal operating conditions.  Section~\ref{sec:conclu} draws the main conclusions and outlines future work directions.

\section{Modeling}
\label{sec:model}


\subsection{Stochastic Differential-Algebraic Equations}
\label{sec:sdae}

We model the power system as a set of differential-algebraic equations:
\begin{equation}
  \label{eq:dae}
  \begin{aligned}
    \dot{\bfg x} &= \bfg f(\bfg x, \bfg y, \bfg \eta) \, , \\ 
    \bfg 0 &= \bfg g(\bfg x, \bfg y, \bfg \eta) \, ,
  \end{aligned}
\end{equation}
where $\bfg f: \mathbb{R}^{l+m+n} \bfg \mapsto \mathbb{R}^{l}$, and $\bfg g: \mathbb{R}^{l+m+n} \bfg \mapsto \mathbb{R}^{m}$ are the differential and algebraic equations; $\bfg x \in \mathbb{R}^{l}$ is the vector of state variables; $\bfg y \in \mathbb{R}^{m}$ is the vector of algebraic variables; $\bfg \eta \in \mathbb{R}^{n}$ represents stochastic processes.  These equations represent the transmission system, synchronous machines, converters, and controllers.  For GFM controllers we utilize two setups, namely a droop control (model REGFM\_A1 \cite{du2023model}) and GFM VSM (model REGFM\_B1 \cite{du2024virtual}).

Stochastic processes (e.g., wind speed) $\bfg \eta$ are modeled as the following set of stochastic differential equations (SDEs):
\begin{equation}
  \label{eq:sde}
  \dot{\bfg \eta} = \bfg a(\bfg \eta) + \bfg b(\bfg \eta) \odot \bfg \xi \, ,
\end{equation}
where ${\bfg a}: \mathbb{R}^{n} \mapsto \mathbb{R}^{n}$, and $\bfg b : \mathbb{R}^{n} \mapsto \mathbb{R}^{n}$ are the vectors of so-called \textit{drift}, and \textit{diffusion} terms, respectively; $\odot$ represents the element-wise product of two vectors; and $\xi(t)$ is the time derivative of the \Wt{} $\bfg W(t)$:
\begin{equation}
  \label{eq:xi}
  \bfg \xi(t) \, dt = d \bfg W(t) \, ,
\end{equation}
where $\bfg W \in \mathbb{R}^{n}$ is a vector of uncorrelated standard \Wt{}es i.e, the elements of $\bfg W(t)$, say $W_i(t)$, $i=1, \dots, n$, are uncorrelated.

Equations \eqref{eq:sde} are SDEs that can be formally integrated to obtain:
\begin{equation} 
  \label{eq:sde_int} 
  \bfg \eta(t) = \bfg \eta(t_{o}) + \int_{t_{o}}^{t} \bfg a(\bfg \eta(s))dt + \int_{t_{o}}^{t} \bfg b(\bfg \eta(s)) \odot d{\bfg W}(s) \, .
\end{equation}
In \eqref{eq:sde_int}, the drift term can be integrated with respect to time along with \eqref{eq:dae} with any conventional numerical integration scheme, for example, trapezoidal method.  On the other hand, the diffusion term has to be treated differently as it depends on the \Wt{}es $d\bfg W(t)$, the increments of which are stochastic and (possibly) unbounded.  In this work, the stochastic integral is usually represented as an \Ito{} integral, and is integrated using \Ito{} calculus \cite{Kloeden:1999, Oksendal:2003}.


\section{Case Study}
\label{sec:case}

We employ the IEEE 9-bus power system to validate the effectiveness of different technologies and control strategies on frequency quality.  Numerous realistic scenarios are considered with extensive time domain simulations performed using Dome \cite{6672387}.  The simulations cover both normal and abnormal system conditions and are evaluated based on real-world frequency quality metrics such as RoCoF, Zenith, $\sigma_{f}$ and minutes outside $\pm$100 mHz.  Additionally, given that not all transmission system operators (TSOs) employ an AGC and the fact that even when AGC is employed different resources operate under it, all scenarios are run with and without AGC similar to \cite{10849615}.  This is a useful comparison, in particular, for TSOs not currently employing an AGC but could consider it in the future such as EirGrid/SONI in Ireland/Northern Ireland and NESO in Great Britain (GB).

\begin{table*}[t!]
  \centering
  \caption{Summary of scenario description.}
  \label{tab:description}
  \resizebox{1.0\linewidth}{!}{
  \begin{tabular}{ccccccccccccc}
    \hline
    Scenario  & $\rm GFL \; Wind$ & $\rm APC$ & $\rm fdb_{wind}$ & $\rm fdb_{conv}$ & $\rm GFL \; BESS$ & $\rm fdb_{bess}$  & $\rm AGC$ & $\rm Load/Wind$ & $\rm Load$ & $\rm Wind$ & $\rm Condenser$ & $\rm GFM \; BESS$  \\
     & Generation& & (mHz) & (mHz) & &(mHz) & conv/wind/GFM & Ramps& Noise& Noise & & VSM/Droop\\
    \hline
    Without AGC \\
    \hline
     1 - Conventional  & No & --  & -- & $\pm$ 15 & No & -- & -- & Load & Yes & -- & -- & -- \\
    2 - GFL Wind $\pm$200 mHz & Yes & Off  & $\pm$ 200 & $\pm$ 15 & No & -- & conv. & Both & Yes & Gaussian & -- & -- \\
    3 - GFL Wind $\pm$15 mHz  & Yes & On  & $\pm$ 15 & $\pm$ 15 & No & -- & conv. & Both & Yes & Gaussian & -- & -- \\
    4 - GFL Wind \& Condenser  & Yes & On  & $\pm$ 15 & $\pm$ 15 & No & -- & conv. & Both & Yes & Gaussian & \OK{} & -- \\
    5 - GFL BESS $\pm$200 mHz  & Yes & On  & $\pm$ 15 & $\pm$ 15 & Yes & $\pm$ 200 & -- & Both & Yes & Gaussian & -- & -- \\
    6 - GFL BESS $\pm$15 mHz  & Yes & On  & $\pm$ 15 & $\pm$ 15 & Yes & $\pm$ 15 & -- & Both & Yes & Gaussian & -- & -- \\
    7 - GFL BESS \& Condenser  & Yes & On  & $\pm$ 15 & $\pm$ 15 & Yes & $\pm$ 15 & -- & Both & Yes & Gaussian & \OK{} & -- \\
    8 - GFM VSM  & No & --  & -- & $\pm$ 15 & No & -- & conv. & Load & Yes & -- & -- & VSM \\
    9 - GFM Droop  & No & --  & -- & $\pm$ 15 & No & -- & conv. & Load & Yes & -- & -- & Droop \\
    10 - GFL Wind \& GFM VSM  & Yes & On  & $\pm$ 15 & $\pm$ 15 & No & -- & conv. & Both & Yes & Gaussian & -- & VSM \\
    11 - GFL Wind \& GFM Droop  & Yes & On  & $\pm$ 15 & $\pm$ 15 & No & -- & conv. & Both & Yes & Gaussian & -- & Droop \\
    \hline
    With AGC \\
    \hline
    1 - Conventional  & No & --  & -- & $\pm$ 15 & No & -- & conv. & Load & Yes & -- & -- & -- \\
    2 - GFL Wind $\pm$200 mHz  & Yes & Off  & $\pm$ 200 & $\pm$ 15 & No & -- & conv. \& wind & Both & Yes & Gaussian & -- & -- \\
    3 - GFL Wind $\pm$15 mHz  & Yes & On  & $\pm$ 15 & $\pm$ 15 & No & -- & conv. \& wind & Both & Yes & Gaussian & -- & -- \\
    4 - GFL Wind \& Condenser  & Yes & On  & $\pm$ 15 & $\pm$ 15 & No & -- & conv. \& wind & Both & Yes & Gaussian & \OK{} & -- \\
    5 - GFL BESS $\pm$200 mHz  & Yes & On  & $\pm$ 15 & $\pm$ 15 & Yes & $\pm$ 200 & conv. & Both & Yes & Gaussian & -- & -- \\
    6 - GFL BESS $\pm$15 mHz  & Yes & On  & $\pm$ 15 & $\pm$ 15 & Yes & $\pm$ 15 & conv. & Both & Yes & Gaussian & -- & -- \\
    7 - GFL BESS \& Condenser  & Yes & On  & $\pm$ 15 & $\pm$ 15 & Yes & $\pm$ 15 & conv. & Both & Yes & Gaussian & \OK{} & -- \\
    8 - GFM VSM  & No & --  & -- & $\pm$ 15 & No & -- & conv. \& GFM & Load & Yes & -- & -- & VSM \\
    9 - GFM Droop  & No & --  & -- & $\pm$ 15 & No & -- & conv. \& GFM & Load & Yes & -- & -- & Droop \\
    10 - GFL Wind \& GFM VSM  & Yes & On  & $\pm$ 15 & $\pm$ 15 & No & -- & conv. \& GFM & Both & Yes & Gaussian & -- & VSM \\
    11 - GFL Wind \& GFM Droop  & Yes & On  & $\pm$ 15 & $\pm$ 15 & No & -- & conv. \& GFM & Both & Yes & Gaussian & -- & Droop \\
    \hline
  \end{tabular}}
\end{table*}

Table~\ref{tab:description} summarizes all the scenarios considered.  A brief description of each of them is also provided below.  

\begin{itemize}
    \item \underline{\textit{1 - Conventional}}: Conventional power systems without wind generation.  Synchronous generators have a $\pm$15 mHz governor dead-band ($\rm fdb_{conv}$).  Represent conventional power systems with synchronous generators operating or not under AGC (setpoints issued every 2s) such as the CE power system (with AGC) and the GB/AIPS (without AGC).
    \item \underline{\textit{2 - GFL Wind $\pm$ 200 mHz}}: Have replaced the conventional synchronous generator connected at bus 3 with a wind power plant which implements a 5-th order doubly-fed
induction generator with voltage, pitch angle,
maximum power point tracking (MPPT)
 controller and PFC with dead-band of $\pm$200 mHz ($\rm fdb_{wind}$).  To be able to provide up and down regulation, we assume wind is operating 20\% below its MPPT (i.e., curtailed).  In terms of AGC, conventional generators always operate under it while wind does (with AGC) or does not participate (without AGC). This represents most of the CE system.
    \item \underline{\textit{3 - GFL Wind $\pm$15 mHz}}: Same as Scenario 2 but with PFC dead-band of wind reduced to $\pm$15 mHz (known as active power control (APC) On in the AIPS).  This  represents the AIPS where APC is enabled regularly.
    \item \underline{\textit{4 - GFL Wind \& Condenser}}: Same as Scenario 3 but now we have also installed a synchronous condenser (at bus 4) modeled as a synchronous generator with zero active power and no turbine governor.   Moreover, the condenser has same inertia constant as the synchronous generator connected at bus 2.  This could represent the AIPS in the near future if the TSOs decide to install an AGC (synchronous condensers are already installed).
    \item \underline{\textit{5 - GFL BESS $\pm$200 mHz}}: Similar to Scenario 3 but now we have also installed a BESS (at bus 4) with PFC dead-band ($\rm fdb_{bess}$) of $\pm$200 mHz (provides primary voltage control as well) \cite{9715680}, and conventional generators operate (with AGC) or not (without AGC) under AGC.  The latter case currently represents the AIPS while the former case could represent the AIPS in the near future if an AGC was to be installed.
    \item \underline{\textit{6 - GFL BESS $\pm$15 mHz}}: Same as Scenario 5 but with BESS having a PFC dead-band of $\pm$15 mHz.  This represent GB where BESS provide PFC services with $\pm$15 mHz dead-band, and in the near future could represent the AIPS as well.
    \item \underline{\textit{7 - GFL BESS \& Condenser}}: Same as Scenario 6 but now we have also installed a synchronous condenser. Could represent both GB and the AIPS.
    \item \underline{\textit{8 - GFM VSM}}: Same as Scenario 1 but now we have replaced a conventional synchronous generator connected at bus 2 with a GFM VSM that operates or not under AGC.  This could represent power systems where GFM has started to be implemented such as Australia \cite{10444667}.
    \item \underline{\textit{9 - GFM Droop}}: Same as Scenario 8 but now we implement GFM droop instead of GFM VSM.  
    \item \underline{\textit{10 - GFL Wind \& GFM VSM}}: This combines Scenario 3 and Scenario 8 where we have replaced two synchronous generators with a wind power plant (connected at bus 3) with PFC dead-band of $\pm$15 mHz and a GFM VSM (connected at bus 2), respectively, where the latter is or not under AGC.
    \item \underline{\textit{11 - GFL Wind \& GFM Droop}}: This is same as Scenario 10 but instead of GFM VSM we implement GFM droop.
\end{itemize}

\subsection{Abnormal Conditions}
\label{sub:abnormal}

One of the main concerns for TSOs and, in particular, for those that operate relatively small islanded power systems, is how to deal with frequency stability or short-term aspects of frequency quality such as keeping RoCoF, Nadir and Zenith within predefined limits following large contingencies/imbalances \cite{10688904}.  In this context, here we consider the loss of load connected at bus 6.  Note that while this represents a significant contingency due to the load loss representing almost 30\% of total load, it is realistic for some power systems such as the AIPS.  As over-frequency is an increasing concern for TSOs, this case study also aims at providing recommendations on how TSOs may deal with this emerging challenge (i.e., TSOs have traditionally been concerned of under-frequency events).

\begin{figure}[thb!]
  \begin{center}
    \resizebox{1.0\linewidth}{!}{\includegraphics{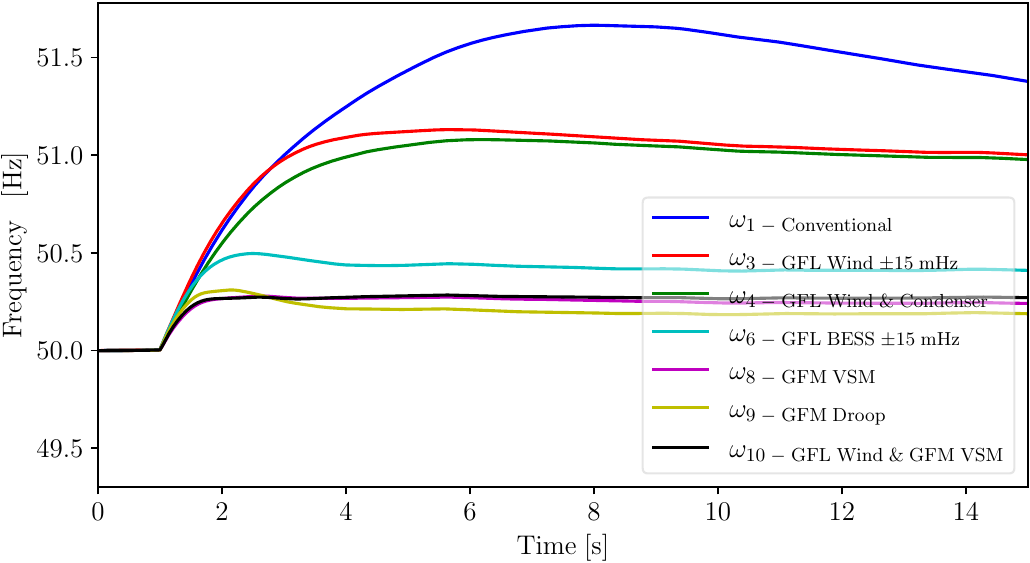}}
    \caption{Frequency trace comparison for contingency event without AGC (loss of load at bus 6).}
    \label{fig:compcont}
  \end{center}
  \vspace{-4mm}
\end{figure}

Figure~\ref{fig:compcont} shows relevant frequency traces whereas Table~\ref{tab:resultscont} presents the results for all scenarios.  The scenarios that include a GFM device (i.e., Scenarios 8, 9, 10 and 11) lead to an excellent frequency response performance.  In contrast, conventional power systems (Scenario 1) suffer larger frequency deviations in terms of Zenith but not RoCoF necessarily due to a lot of inertia.  These can be better seen in Table~\ref{tab:resultscont}.  Note that RoCoF is calculated over a rolling window of 500 ms which is a standard in the industry.

Comparing GFM control strategies, that is, GFM VSM and GFM droop, it appears that GFM VSM leads to a slightly better performance in terms of Zenith and RoCoF (e.g., Zenith of 50.28 and RoCoF of 0.42, Scenario 8, compared to Zenith of 50.31 and RoCoF of 0.51 Hz/s, Scenario 9).  In addition to Zenith, time to Zenith ($T_{\rm zenith}$) and RoCoF metrics, we show in Table~\ref{tab:resultscont} also the time to restore ($T_{\rm restore}$) frequency within defined ranges such as $\pm$200 mHz as is the case in the AIPS.  In contrast to Zenith and RoCoF results, GFM droop outperforms GFM VSM in terms of $T_{\rm restore}$.  In Table \ref{tab:resultscont}, Scenarios are also classified as ``Insecure'' or ``Secure'' if these metrics are outside or inside operational limits for the AIPS system (e.g., Zenith limit of 51 Hz and RoCoF limit of 1 Hz/s).  

Another interesting result is that related to impact of PFC dead-band of wind and BESS.  Specifically, it appears that there is no significant difference on frequency stability if the dead-band of wind PFC is $\pm$200 mHz or $\pm$15 mHz (Scenarios 2 and 3).  For instance, if APC is On (dead-band of $\pm$15 mHz) leads to a RoCoF of 0.72 Hz/s while when APC is Off leads to a RoCoF of 0.73 Hz/s.  This is not the case for BESS (Scenarios 5 and 6) where it can be seen that having a narrow dead-band, in this case $\pm$15 mHz, leads to improved Zenith and RoCoF (i.e., from 0.70 Hz/s to 0.63 Hz/s).  This is considered useful information for TSOs looking into the management of PFC settings to deal with frequency stability such as the GB and Irish TSOs \cite{kerci2025comprehensive}.

The cases with and without AGC lead to very similar results except time to restore frequency within the $\pm$200 mHz range ($T_{\rm restore}$).  This is to be expected considering AGC is a slow controller compared to PFC, in particular, PFC of IBRs such as BESS.  This is another useful information for TSOs considering installing an AGC.  In other words, if frequency restoration is not an issue for TSOs, then an AGC may not be needed at all from a short-term frequency quality or frequency stability perspective.  The impact of AGC on long-term frequency performance is discussed in the next section.

Looking at the contingency results, we can conclude that while GFM appears to significantly improve frequency stability, a combination of GFL devices such as wind and BESS and synchronous condensers (i.e., Scenarios 6 and 7) lead to similar and acceptable frequency performance (e.g., Zenith below 50.5 Hz and RoCoF below 0.60 Hz/s).  For this reason, the choice between GFM or GFL plus synchronous condensers might be more economical or driven by other issues such as system strength rather than for short-term frequency quality.  However, our opinion, which is based on the results of this paper, is that TSOs operating small islanded systems should consider integration of GFM to increase robustness of their systems against large contingencies.

\begin{table}[h!]
  \centering
  \caption{Results for contingency events.} 
  \label{tab:resultscont}
  \resizebox{1.0\linewidth}{!}{
  \begin{tabular}{lccccc}
    \hline
    Scenario  & Zenith  & $T_{\rm zenith}$ & RoCoF & $T_{\rm restore}$ & Security\\
     & (Hz) & (s) & (Hz/s) & (s)\\
     \hline
    Without AGC \\
    \hline
    1 - Conventional & 51.66 & 6.97 & 0.66 & No rest. & Insecure
    \\
    2 - GFL Wind $\pm$200 mHz & 51.25 & 4.79 & 0.73 & 49.0 & Insecure  \\
    3 - GFL Wind $\pm$15 mHz & 51.13 & 4.62 & 0.72 & 45.52 & Insecure  \\
    4 - GFL Wind \& Condenser & 51.08 & 5.19 & 0.59  & 45.47 & Insecure   \\
    5 - GFL BESS $\pm$200 mHz & 50.60 & 1.58 & 0.70  & No rest. & Insecure   \\
    6 - GFL BESS $\pm$15 mHz  & 50.49 & 1.50 & 0.63  & No rest. & Insecure   \\
    7 - GFL BESS \& Condenser & 50.47 & 1.73 & 0.53  & No rest. & Insecure   \\
    8 - GFM VSM & 50.28 & 1.63 & 0.42  & 23.47 & Secure   \\
    9 - GFM Droop & 50.31 & 1.16 & 0.51 & 3.44 & Secure   \\
    10 - GFL Wind \& GFM VSM & 50.28 & 4.58 & 0.44 & 93.66  & Secure  \\
    11 - GFL Wind \& GFM Droop & 50.30 & 0.89 & 0.54 & 21.62 & Secure   \\
    \hline
    With AGC \\
    \hline
        1 - Conventional & 51.62 & 6.76 & 0.66 & 17.84 & Insecure
    \\
    2 - GFL Wind $\pm$200 mHz  & 51.23 & 4.59 & 0.73 & 30.02 & Insecure  \\
    3 - GFL Wind $\pm$15 mHz & 51.10 & 4.55 & 0.72 & 27.90 & Insecure  \\
    4 - GFL Wind \& Condenser & 51.05 & 5.0 & 0.59  & 28.05 & Insecure   \\
    5 - GFL BESS $\pm$200 mHz  & 50.60 & 1.60 & 0.70  & 66.08 & Secure   \\
    6 - GFL BESS $\pm$15 mHz & 50.49 & 1.50 & 0.63  & 39.80 & Secure   \\
    7 - GFL BESS \& Condenser & 50.47 & 1.74 & 0.53  & 39.83 & Secure   \\
    8 - GFM VSM & 50.28 & 1.63 & 0.42  & 19.61 & Secure   \\
    9 - GFM Droop & 50.31 & 1.16 & 0.51 & 3.24 & Secure   \\
    10 - GFL Wind \& GFM VSM & 50.28 & 4.56 & 0.44 & 28.20  & Secure  \\
    11 - GFL Wind \& GFM Droop & 50.30 & 0.89 & 0.54 & 12.07 & Secure   \\
    \hline
  \end{tabular}}
\end{table}

\subsection{Normal Conditions}
\label{sub:normal}

Long-term frequency quality which is mainly related to normal operating conditions is more often than not overlooked in the literature despite TSOs dedicating of lot of time and effort to keep it within defined standards \cite{entsoe}.  In this context, this section aims to fill this gap.  To do so, we perform long-term dynamic stochastic simulations namely 24h, and quantify long-term frequency quality using different metrics such as $\sigma_{f}$, asymmetry of $\sigma_{f}$ namely $\Delta \sigma_{f}$, and minutes outside $\pm$100 mHz \cite{10849615}.  In particular, we have introduced realistic wind and demand ramps and noise.  The detailed model of stochastic disturbances is given in \cite{10849615}.

\begin{figure}[thb!]
  \begin{center}
    \resizebox{1.0\linewidth}{!}{\includegraphics{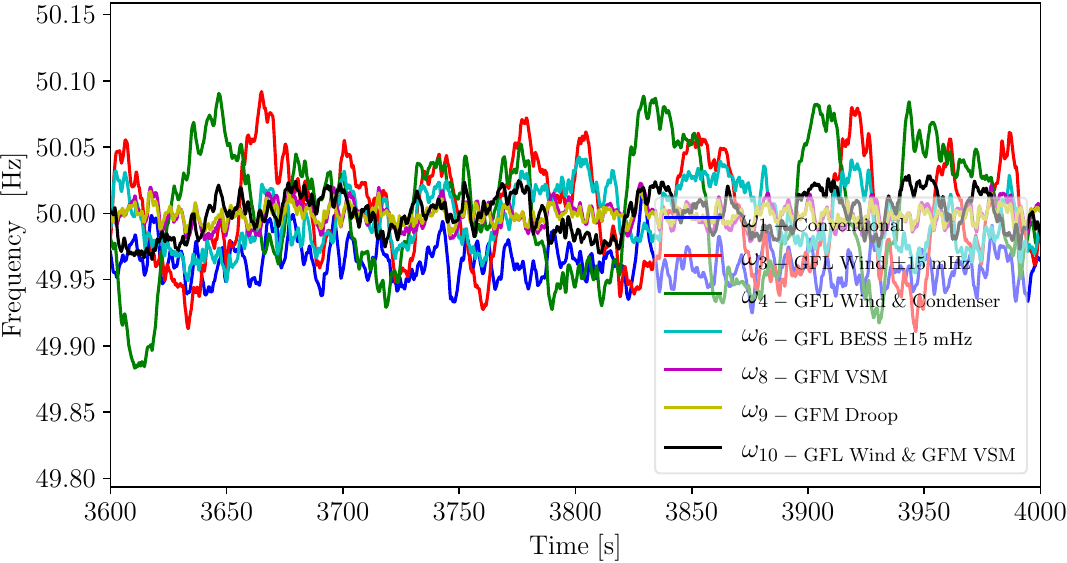}}
    \caption{Frequency trace comparison for normal conditions without AGC.}
    \label{fig:comp}
  \end{center}
  \vspace{-4mm}
\end{figure}

Figure~\ref{fig:comp} illustrates the results for same scenarios as in the previous section using a random window in the simulation, whereas Table~\ref{tab:results} provides detailed results for all scenarios.   
Similar to contingency events, Figure~\ref{fig:comp} and Table~\ref{tab:results} suggest that, overall, GFM significantly also improves long-term frequency quality.  For instance, Scenario 8, 9, 10 and 11 that include GFM devices show very small $\sigma_{f}$, $\Delta \sigma_{f}$ (almost by an order of magnitude) and zero minutes outside $\pm$100 mHz range when compared with other relevant scenarios such as Scenario 3, 4, 5, 6 and 7 that involve GFL devices such as wind and BESS and synchronous condensers.  As a matter of fact, and for comparison, Scenarios 3 to 7 show more or less same $\sigma_{f}$ as the Nordic, AIPS and CE systems \cite{entsoe}.

\begin{table*}[h!]
  \centering
  \caption{Results for normal conditions.} 
  \label{tab:results}
  \begin{tabular}{lccccccccc}
    \hline
    Scenario  & Mean  & $\sigma_{f}$ & $\sigma_{f-}$ & $\sigma_{f+}$ & $\Delta \sigma_{f}$ & $\rm Outside \pm 100 mHz$ & $< 49.9 \; \rm Hz$ & $> 50.1 \; \rm Hz$\\
     & (Hz) & (Hz) & (Hz) & (Hz) & (Hz) & (mins) & (mins) & (mins) \\
    \hline
    Without AGC \\
    \hline
    1 - Conventional & 49.99 & 0.02490 & 0.02572 & 0.0240 & 0.0017 & 0 & 0 & 0 \\
    2 - GFL Wind $\pm$200 mHz  & 50.00 & 0.03715 & 0.03709 & 0.03821 & 0.00011 & 5.83 & 3.47 & 2.35 \\
    3 - GFL Wind $\pm$15 mHz & 50.00 & 0.03922 & 0.03939 & 0.03905 & 0.00034 & 10.46 & 6.60 & 3.85 \\
    4 - GFL Wind \& Condenser & 50.00 & 0.03731 & 0.03772 & 0.03690 & 0.00081 & 8.93 & 6.78 & 2.15 \\
    5 - GFL BESS $\pm$200 mHz  & 50.01 & 0.0511 & 0.0456 & 0.0539 & 0.0083 & 13.73 & 7.17 & 6.55 \\
    6 - GFL BESS $\pm$15 mHz & 50.00 & 0.03204 & 0.02693 & 0.03499 & 0.00805 & 0 & 0 & 0 \\
    7 - GFL BESS \& Condenser & 50.00 & 0.03169 & 0.02655 & 0.03465 & 0.00809 & 0 & 0 & 0 \\
    8 - GFM VSM & 50.00 & 0.009478 & 0.009499 & 0.009459 & 0.000040 & 0 & 0 & 0 \\
    9 - GFM Droop & 50.00 & 0.00627 & 0.00628 & 0.00627 & 0.000016 & 0 & 0 & 0 \\
    10 - GFL Wind \& GFM VSM & 50.00 & 0.01424 & 0.01420 & 0.01428 & 0.000082 & 0 & 0 & 0 \\
    11 - GFL Wind \& GFM Droop & 50.00 & 0.01005 & 0.01006 & 0.01004 & 0.000018 & 0 & 0 & 0 \\
    \hline
    With AGC \\
    \hline
    1 - Conventional & 50.00 & 0.02165 & 0.02164 & 0.02165 & 0.000013 & 0 & 0 & 0 \\
    2 - GFL Wind $\pm$200 mHz & 50.00 & 0.03545 & 0.03542 & 0.03549 & 0.000061 & 5.01 & 3.32 & 1.69 \\
    3 - GFL Wind $\pm$15 mHz & 50.00 & 0.03546 & 0.03565 & 0.0352 & 0.00036 & 7.94 & 5.91 & 2.03 \\
    4 - GFL Wind \& Condenser & 50.00 & 0.03421 & 0.03436 & 0.03407 & 0.00028 & 6.32 & 4.54 & 1.78 \\
    5 - GFL BESS $\pm$200 mHz & 50.00 & 0.0395 & 0.0397 & 0.0393 & 0.00039 & 12.23 & 7.75 & 4.48 \\
    6 - GFL BESS $\pm$15 mHz & 50.00 & 0.0221 & 0.0222 & 0.0220 & 0.00019 & 0 & 0 & 0 \\
    7 - GFL BESS \& Condenser & 50.00 & 0.02168 & 0.02179 & 0.02158 & 0.00021 & 0 & 0 & 0 \\
    8 - GFM VSM  & 50.00 & 0.00945 & 0.00948 & 0.00943 & 0.000047 & 0 & 0 & 0 \\
    9 - GFM Droop & 50.00 & 0.00626 & 0.00627 & 0.00625 & 0.000017 & 0 & 0 & 0 \\
    10 - GFL Wind \& GFM VSM & 50.00 & 0.01398 & 0.01393 & 0.0140 & 0.00091 & 0 & 0 & 0 \\
    11 - GFL Wind \& GFM Droop & 50.00 & 0.00985 & 0.00984 & 0.00986 & 0.000021 & 0 & 0 & 0 \\
   \hline
  \end{tabular}
\end{table*}

Another interesting result is that GFM droop (e.g., Scenario 9) appears to slightly outperform GFM VSM (e.g., Scenario 8) in terms of $\sigma_{f}$ and $\Delta \sigma_{f}$.  This is different to the contingency results where GFM VSM slightly outperformed GFM droop in terms of Zenith and RoCoF.  These results suggest that TSOs facing challenges with frequency stability may consider the installation of GFM VSM devices while others that suffer from poor long-term frequency quality may consider GFM droop devices, assuming the economics are same or very similar.  Note that we are aware that with proper tuning different GFM control strategies, and even GFL devices such as BESS, may lead to same or similar results but this is outside of the scope of this paper (i.e., we have done little to no tuning compared to published parameters in \cite{du2023model, du2024virtual}).

While the cases with AGC generally lead to lower $\sigma_{f}$, $\Delta \sigma_{f}$ and minutes outside $\pm$100 mHz range, the differences are not significant.  In fact, there are many scenarios without AGC, for example, all GFM Scenarios (8-11) and Scenarios with BESS devices with $\pm$15 mHz PFC dead-band (Scenarios 6 and 7) that lead to zero minutes outside $\pm$100 mHz range.  These results really question the importance of AGC in IBR-dominated grids and, in particular, in GFM IBR-dominated grids due to their excellent inertial and PFC capability and, in turn, increased frequency quality.

\section{Concluding Remarks}
\label{sec:conclu}

This paper provides a comprehensive assessment of the impact of conventional and emerging power technologies and control strategies on frequency quality.  The scenarios included in the case study consider GFM and GFL IBRs, as well as synchronous generators and synchronous condensers.

Extensive dynamic simulations and realistic scenarios on the IEEE 9-bus system lead to conclude that GFM significantly outperforms other technologies (e.g., synchronous generators/condensers and GFL wind) in terms of both short and long-term aspects of frequency quality such as dealing with Zenith, RoCoF and $\sigma_{f}$.  Despite this, a combination of these other technologies, for example, GFL wind, BESS and synchronous condensers, lead to lower but acceptable frequency performance compared to GFM.  For this reason, the choice between GFM and GFL plus synchronous condensers may be driven by economics or by other system issues such as system strength \cite{10508461}.

Another insightful result of the paper is that the frequency control role and importance of AGC in GFM IBR-dominated grids becomes less evident due to GFM providing excellent inertial and PFC and, thus, reducing the need for AGC.  But we believe that the question whether AGC is needed at all or not deserves further detailed considerations.  These include: (i) Applying AGC in a real-world power system model and testing its effectiveness under different operating conditions similar to this work; (ii) AGC benefits with regard to enabling cross-border balancing energy exchange and imbalance netting process;  (iii) AGC benefits in terms of helping maintain scheduled power flows between different balancing control areas (tie-line active power control); (iv) AGC benefits in terms of removing or, at least, significantly reducing the need for issuing manual dispatch instructions; (v) AGC support with automatic time error control (in jurisdictions where it is still performed); and last but not least (vi) Role and benefits of AGC in self vs central dispatch systems.  These aspects will be considered in future work.


\begin{thebibliography}{10}
\providecommand{\url}[1]{#1}
\csname url@samestyle\endcsname
\providecommand{\newblock}{\relax}
\providecommand{\bibinfo}[2]{#2}
\providecommand{\BIBentrySTDinterwordspacing}{\spaceskip=0pt\relax}
\providecommand{\BIBentryALTinterwordstretchfactor}{4}
\providecommand{\BIBentryALTinterwordspacing}{\spaceskip=\fontdimen2\font plus
\BIBentryALTinterwordstretchfactor\fontdimen3\font minus
  \fontdimen4\font\relax}
\providecommand{\BIBforeignlanguage}[2]{{%
\expandafter\ifx\csname l@#1\endcsname\relax
\typeout{** WARNING: IEEEtran.bst: No hyphenation pattern has been}%
\typeout{** loaded for the language `#1'. Using the pattern for}%
\typeout{** the default language instead.}%
\else
\language=\csname l@#1\endcsname
\fi
#2}}
\providecommand{\BIBdecl}{\relax}
\BIBdecl

\bibitem{entsoe}
\BIBentryALTinterwordspacing
{ENTSO-E}, ``2023 annual load-frequency control report,'' 2024. [Online].
  Available:
  \url{https://eepublicdownloads.entsoe.eu/clean-documents/SOC%20documents/LFC/ALFC_report_2023.pdf}
\BIBentrySTDinterwordspacing

\bibitem{10253411}
T.~Kërçi \emph{et~al.}, ``Frequency quality in low-inertia power systems,''
  in \emph{2023 IEEE Power \& Energy Society General Meeting (PESGM)}, 2023,
  pp. 1--5.

\bibitem{10688761}
------, ``Emerging challenges of integrating solar {PV in the Ireland and
  Northern Ireland} power systems,'' in \emph{2024 IEEE Power \& Energy Society
  General Meeting (PESGM)}, 2024, pp. 1--5.

\bibitem{8879610}
J.~Matevosyan \emph{et~al.}, ``Grid-forming inverters: Are they the key for
  high renewable penetration?'' \emph{IEEE Power and Energy Magazine}, vol.~17,
  no.~6, pp. 89--98, 2019.

\bibitem{10444681}
B.~Bahrani \emph{et~al.}, ``Grid-forming inverter-based resource research
  landscape: Understanding the key assets for renewable-rich power systems,''
  \emph{IEEE Power and Energy Magazine}, vol.~22, no.~2, pp. 18--29, 2024.

\bibitem{9714816}
Y.~Li \emph{et~al.}, ``Revisiting grid-forming and grid-following inverters: A
  duality theory,'' \emph{IEEE Transactions on Power Systems}, vol.~37, no.~6,
  pp. 4541--4554, 2022.

\bibitem{9513281}
D.~B. Rathnayake \emph{et~al.}, ``Grid forming inverter modeling, control, and
  applications,'' \emph{IEEE Access}, vol.~9, pp. 114\,781--114\,807, 2021.

\bibitem{11008671}
S.~L. Skogen \emph{et~al.}, ``Assessing oscillatory stability with dominant
  grid-forming power systems for active power imbalances,'' \emph{IEEE Open
  Access Journal of Power and Energy}, vol.~12, pp. 318--329, 2025.

\bibitem{jalali2023impact}
A.~Jalali \emph{et~al.}, ``Impact of grid-forming inverters on frequency
  control of a grid with high share of inverter-based resources,'' in
  \emph{Proceedings of the CAIRNS International Symposium, Cairns, Australia},
  2023, pp. 4--7.

\bibitem{iet}
S.~Rogalla \emph{et~al.}, ``Grid-forming converters in interconnected power
  systems: Requirements, testing aspects, and system impact,'' \emph{IET
  Renewable Power Generation}, vol.~18, no.~15, pp. 3053--3066, 2024.

\bibitem{du2023model}
W.~Du \emph{et~al.}, ``Model specification of droop-controlled, grid-forming
  inverters {(REGFM\_A1)},'' Pacific Northwest National Laboratory (PNNL),
  Richland, WA (United States), Tech. Rep., 2023.

\bibitem{du2024virtual}
------, ``Virtual synchronous machine grid-forming inverter model specification
  {(REGFM\_B1)},'' National Renewable Energy Laboratory (NREL), Golden, CO
  (United States), Tech. Rep., 2024.

\bibitem{Kloeden:1999}
E.~Kl\"{o}den \emph{et~al.}, \emph{Numerical Solution of Stochastic
  Differential Equations}, 3rd~ed.\hskip 1em plus 0.5em minus 0.4em\relax
  Springer, 1999.

\bibitem{Oksendal:2003}
B.~\O{}ksendal, \emph{Stochastic Differential Equations: An Introduction with
  Applications}.\hskip 1em plus 0.5em minus 0.4em\relax New York, 6th ed.:
  Springer, 2003.

\bibitem{6672387}
F.~Milano, ``A {Python}-based software tool for power system analysis,'' in
  \emph{2013 IEEE Power \& Energy Society General Meeting}, 2013, pp. 1--5.

\bibitem{10849615}
T.~Kërçi \emph{et~al.}, ``Asymmetry of frequency distribution in power
  systems: Sources, estimation, impact and control,'' \emph{IEEE Open Access
  J.~of Power and Energy}, vol.~12, pp. 135--145, 2025.

\bibitem{9715680}
J.~McMahon \emph{et~al.}, ``Combining flexible loads with energy storage
  systems to provide frequency control,'' in \emph{2021 IEEE PES Innovative
  Smart Grid Technologies - Asia (ISGT Asia)}, 2021, pp. 1--5.

\bibitem{10444667}
B.~Badrzadeh \emph{et~al.}, ``Grid-forming inverters: Project demonstrations
  and pilots,'' \emph{IEEE Power and Energy Magazine}, vol.~22, no.~2, pp.
  66--77, 2024.

\bibitem{10688904}
M.~Hurtado \emph{et~al.}, ``Stability assessment of low-inertia power systems:
  A system operator perspective,'' in \emph{2024 IEEE Power \& Energy Society
  General Meeting (PESGM)}, 2024, pp. 1--5.

\bibitem{kerci2025comprehensive}
T.~Kërçi \emph{et~al.}, ``A comprehensive approach to evaluate frequency
  control strength of power systems,'' \emph{arXiv preprint arXiv:2509.00548},
  2025.

\bibitem{10508461}
H.~Xin \emph{et~al.}, ``How many grid-forming converters do we need? {A}
  perspective from small signal stability and power grid strength,'' \emph{IEEE
  Transactions on Power Systems}, vol.~40, no.~1, pp. 623--635, 2025.

\end{thebibliography}


\end{document}